\documentclass[twocolumn,aps,prd,groupedaddress,nofootinbib,superscriptaddress,floatfix,amsmath,amssymb,preprintnumbers]{revtex4-1}

\usepackage{CJK}

\usepackage{graphicx}
 \usepackage{epstopdf}
\usepackage{amsmath}
\usepackage{dcolumn}
\usepackage{bm}
\usepackage{slashed}
\usepackage{siunitx}
\usepackage{ulem,xpatch}
\usepackage{hyperref}
\usepackage[mathlines]{lineno}
\usepackage{comment}
\usepackage{soul}
\usepackage{xcolor}
\usepackage{xfrac}
\hypersetup{colorlinks, linkcolor = [rgb]{0,0.0,0.5}, citecolor = [rgb]{0,0.0,0.5}, urlcolor = [rgb]{0,0.0,0.5}}

\newcommand{\mbf}[1]{\mathbf{#1}}
\newcommand{\req}[1]{(\ref{#1})}
\newcommand{\half}{{\textstyle{\frac{1}{2}}}}
\newcommand{\fourth}{{\textstyle{\frac{1}{4}}}}

\def\lb{\label}

\begin{document}

\title{Longitudinal dynamics and chiral symmetry breaking in holographic light-front QCD}

\newcommand*{\COSTARICA}{Laboratorio de F\'isica Te\'orica y Computacional, Universidad de Costa Rica, 11501 San Jos\'e, Costa Rica}\affiliation{\COSTARICA}
\newcommand*{\SLAC}{SLAC National Accelerator Laboratory, Stanford University, Stanford, CA 94309, USA}\affiliation{\SLAC}

\author{Guy~F.~de~T\'eramond}\email{guy.deteramond@ucr.ac.cr}\affiliation{\COSTARICA}
\author{Stanley~J.~Brodsky}\email{sjbth@slac.stanford.edu}\affiliation{\SLAC}

\date{\today}

\begin{abstract}

The breaking of chiral symmetry in holographic light-front QCD is encoded in its longitudinal dynamics, with its chiral limit protected by the superconformal algebraic structure which governs its transverse dynamics.  The  scale in the longitudinal light-front Hamiltonian determines the confinement strength in this direction: It is also responsible for most of the light meson ground state mass consistent with the Gell-Mann-Oakes-Renner constraint. Longitudinal confinement and the breaking of chiral symmetry are found to be different manifestations of the same underlying dynamics as found in the  't Hooft large-$N_C$ QCD (1 + 1) model.

 \end{abstract}

\maketitle

\section{Introduction \lb{introduction}}

In spite of the important progress of Euclidean lattice gauge theory, a basic understanding of the mechanism of color confinement and its relation to chiral symmetry breaking in QCD, two fundamental phenomena of strong interactions,  has remained an unsolved problem.  Recent developments based on superconformal quantum mechanics~\cite{deAlfaro:1976je, Fubini:1984hf} in light-front quantization~\cite{Dirac:1949cp} and its holographic embedding on a higher dimensional gravity theory~\cite{Maldacena:1997re} (gauge/gravity correspondence) have led to new analytic insights into the structure of hadrons and their dynamics~\cite{deTeramond:2008ht, Brodsky:2013ar, deTeramond:2014asa, Dosch:2015nwa, Brodsky:2014yha, Brodsky:2020ajy}. This new approach to nonperturbative QCD dynamics, holographic light-front QCD, leads to effective semi-classical relativistic bound-state equations for arbitrary spin~\cite{deTeramond:2013it}, and it incorporates fundamental properties which are not apparent from the QCD Lagrangian, such as the emergence of the hadron mass scale, the prediction of a massless pion in the chiral limit, and the remarkable connections between meson, baryon and tetraquark spectroscopy across the full hadron spectrum~\cite{Dosch:2015bca, Dosch:2016zdv, Nielsen:2018uyn, Nielsen:2018ytt}.  Phenomenological extensions of the holographic QCD approach  provide nontrivial connections between the dynamics of form factors and polarized and unpolarized quark distributions with pre-QCD nonperturbative approaches such as Regge theory and the Veneziano model~\cite{Sufian:2016hwn, deTeramond:2018ecg, Liu:2019vsn}, and also describe the running of the QCD coupling $\alpha_s(Q^2)$ in the nonperturbative domain~\cite{Brodsky:2010ur, Deur:2014qfa}

In this article we examine the effect of longitudinal light-front (LF) dynamics for the computation of hadron masses, confinement, and chiral symmetry breaking motivated by the previous work in Refs.~\cite{tHooft:1974pnl, Bergknoff:1976xr, Zhitnitsky:1985um, Hornbostel:1988fb, Mo:1992sv,  Brodsky:2008pg, Chabysheva:2012fe, Gutsche:2012ez, Trawinski:2014msa, Li:2015zda, Sheckler:2020fbt}. Although the semiclassical model approximation to LF quantized QCD determines the confinement potential in the LF transverse coordinates  in the  zero quark mass limit~\cite{Brodsky:2020ajy}, an extension is required which incorporates color-confining LF longitudinal dynamics for non-zero quark masses~\cite{AdSQCD}. This extension of holographic LF QCD (HLFQCD) should preserve its successful predictions and incorporate chiral symmetry breaking, while restoring 3-dimensional rotational invariance in the heavy-quark limit.

The main contents of this article are  as follows. In Sec. \ref{LD} we briefly review the extension of the holographic light-front framework to include quark masses by combining the longitudinal dynamics with the HLFQCD transverse dynamics  by following~\cite{Chabysheva:2012fe, Li:2015zda}. In Sec. \ref{U} we recall how the transverse confinement potentials for mesons, baryons and tetraquarks are determined by the underlying conformal symmetry of light-front QCD and its embedding in AdS space~\cite{Brodsky:2020ajy}. We further examine in this section two models of longitudinal confinement: The  't Hooft large-$N_C$ model in one-space and one time-time dimension, as the prototypical example of a simple model derived from QCD, which allows the computation of a meson spectrum while incorporating chiral symmetry breaking. We also examine in this section the longitudinal potential model introduced in \cite{Li:2015zda} which generates a complete basis function and leads to  important constraints in the large quark mass non-relativist limit. These aspects, together with the HLFQCD constraints, are used in Sec. \ref{LFEM} to extend our previous superconformal holographic results~\cite{Brodsky:2014yha, Brodsky:2020ajy} to include longitudinal dynamics and chiral symmetry breaking in the semiclassical Hamiltonian light-front wave equations. Numerical results  are presented in Sec. \ref{NR},  including a computation of meson masses and distribution amplitudes. Some final comments are included in Sec. \ref{CaO}.

\section{Longitudinal and transverse dynamics in HLFQCD \lb{LD}}

We start with the light-front invariant QCD Hamiltonian
$H^{LF} = P^2 = P^+ P^- - \mathbf{P}^2_\perp$, $P^2 = P_\mu P^\mu = M^2$,
\begin{align} \lb{HLF}
H^{LF} \vert \psi \rangle  = M^2  \vert \psi \rangle,
\end{align}
and compute $M^2$  from  the hadronic matrix element $\langle \psi(P’) \vert H^{LF} \vert \psi(P)\rangle = M^2  \langle \psi(P’) \vert  \psi(P)\rangle$, expanding the hadronic states in terms of their Fock components~\cite{deTeramond:2008ht, Brodsky:2014yha}. For the simplest hadronic bound-state, a two parton state with quark masses $m_q$ and $m_{\bar q}$,
\begin{multline}  \lb{M2k}
M^2 = \\   \int_0^1 \! d x \! \int   \frac{d^2 \mbf{k}_\perp}{16 \pi^3}   
\Big(\frac{\mbf{k}_\perp^2 }{x(1-x)} + \frac{m_q^2}{x} + \frac{ m_{\bar q}^2}{(1-x)}  \Big)
 \left\vert \psi (x, \mbf{k}_\perp) \right \vert^2  \\
+   \int_0^1 \! d x  \! \int   \frac{d^2 \mbf{k}_\perp}{16 \pi^3}   \, U(x, \mbf{k}_\perp) 
 \left\vert \psi (x, \mbf{k}_\perp) \right \vert^2 ,
\end{multline}
with normalization
\begin{align} \lb{Nk}
\int_0^1 \!  d x  \! \int  \frac{d^2 \mbf{k}_\perp}{16 \pi^3}   \,  \left\vert \psi (x, \mbf{k}_\perp) \right \vert^2 = 1.
\end{align}
The variable $x$  is the LF longitudinal momentum fraction  $x = k^+/P^+$ and  $\mbf{k}_\perp$ is  the relative transverse momentum. Taking the Fourier transform, Eq. \req{M2k} can be expressed in terms of the transverse impact variable $\mbf{b}_\perp$, conjugate to $\mbf{k}_\perp$, 
\begin{multline}  \lb{M2b}
M^2  =  \\
\int_0^1 \! d x  \! \int  \! d^2 \mbf{b}_\perp 
  \psi^*(x, \mbf{b}_\perp)
  \Big( \frac{- \mbf{\nabla}_{\mbf{b}_ \perp}^2}{x(1-x)} +  \frac{m_q^2}{x} + \frac{ m_{\bar q}^2}{(1-x)}\Big)
  \psi(x, \mbf{b}_\perp) \\
   +  \int_0^1 \! d x  \! \int   d^2 \mbf{b}_\perp  \,
  \psi^*(x, \mbf{b}_\perp)  U(x, \mbf{b}_\perp)   \psi(x, \mbf{b}_\perp) ,
 \end{multline}
with normalization
\begin{align} \lb{Nb}
 \int_0^1 \!  d x \! \int   d^2 \mbf{b}_\perp  \, \left\vert\psi(x, \mbf{b}_\perp)\right\vert^2 = 1.
\end{align}

As noted in Refs.~\cite{Chabysheva:2012fe, Li:2015zda},  the LF Hamiltonian equation~\req{HLF} leads to a longitudinal and a transverse light-front wave equation in the approximation where transverse and longitudinal dynamics are separated. To show this, we  introduce the invariant LF transverse variable $\zeta^2  = x(1-x)  \mbf{b}^2_\perp$, which is identified with the holographic AdS variable $z$~\cite{Brodsky:2014yha}, and factor out the longitudinal, transverse and orbital dependence of the LF wave function~\cite{deTeramond:2008ht, Brodsky:2014yha}
\begin{align} \lb{psi}
\psi(x, \zeta, \varphi) =  e^{i L \varphi}  X(x) \frac{\phi(\zeta)}{\sqrt{2 \pi \zeta}}, 
\end{align}
in the approximation where the effective LF confinement potential  is written as the sum of longitudinal and transverse components~\cite{Chabysheva:2012fe, Li:2015zda}
\begin{align} \lb{Uxb}
 U(x, \zeta) =  U_\perp(\zeta) +  U_\parallel(x) .
\end{align}
The LF wave function normalization \req{Nb} is equivalent to
\begin{align} 
 \int_0^\infty \!  d \zeta \, \phi^2(\zeta)  &= 1, \\
 \int_0^1 dx \, \frac{X^2(x)}{x(1-x)} &=1. \lb{NX}
 \end{align}

Following the procedure introduced in~\cite{deTeramond:2008ht}, we obtain for \req{M2b}
\begin{multline} \lb{M2zetax}
M^2  = \\ \int_0^\infty  \! d\zeta \, \phi^*(\zeta) \sqrt{\zeta}
\left( -\frac{d^2}{d\zeta^2} -\frac{1}{\zeta} \frac{d}{d\zeta}
+ \frac{L^2}{\zeta^2} + U_\perp(\zeta) \right)
\frac{\phi(\zeta)}{\sqrt{\zeta}}  \\
+ \int_0^1 \! dx \, \chi^*(x) \left( \frac{m_q^2}{x} + \frac{ m_{\bar q}^2}{(1-x)} + U_\parallel(x)\right) \chi(x) ,
\end{multline}
where  $L$ is the relative LF orbital angular momentum $L \equiv \vert L^z \vert_{max}$ and
$X(x) =  \sqrt{x(1-x)} \, \chi(x)$, with the normalization of $\chi(x)$ determined from \req{NX}
\begin{align} \lb{norm}
 \int_0^1 dx \, \chi^2(x)  = 1.
 \end{align}
In the chiral limit  $X(x) \to  \sqrt{x(1-x)}$~\cite{Brodsky:2006uqa}, and therefore $\chi(x) = 1$ for zero quark masses.
 
Eq. \req{M2zetax} is thus equivalent to the semiclassical 
 LF transverse~\cite{deTeramond:2008ht, Brodsky:2014yha} and longitudinal~\cite{Chabysheva:2012fe, Li:2015zda}  wave equations for mesons 
 \begin{align}\lb{HT}
 \Big(-\frac{d^2}{d\zeta^2} 
- \frac{1 - 4L^2}{4\zeta^2}+ U_\perp(\zeta) \Big)  \phi(\zeta) &= M_\perp^2 \phi(\zeta) ,   \\
 \lb{HL}
\Big( \frac{m_q^2}{x} +  \frac{m_{\bar q}^2}{1-x}  + U_\parallel(x)\Big) \chi(x) &= M_\parallel^2  \, \chi(x),
\end{align}
with   $M^2 = M_\perp ^2 + M_\parallel^2$. Both equations \req{HT} and \req{HL} are relativistic and frame independent. We note that the separation of longitudinal and transverse equations \req{HT} and \req{HL}  is natural in the light-front, since the quark masses are associated with the longitudinal component of the kinetic energy, not the transverse part~\cite{Chabysheva:2012fe}.

As noted in Refs.~\cite{Chabysheva:2012fe, Li:2015zda},  the longitudinal eigenvalue equation \req{HL} for the longitudinal mass $M_\parallel^2$ can be combined with the holographic LF transverse equation \req{HT} for $M_\perp^2$ to incorporate massive quarks. The longitudinal mass $M_\parallel^2$  appears as a separation constant in the transverse equation \req{HT}, namely $M_\perp^2 \! \to  \! M^2 \! - M_\parallel^2$~\cite{Chabysheva:2012fe}. As a result, the structure of the superconformal equation in the transverse direction is not modified, even by heavy quark masses, as long as transverse and longitudinal dynamics can be separated.

\subsection{HLFQCD equations for baryons}

The semiclassical approximation to the light-front Hamiltonian for baryons is given by a system of two coupled linear differential equations for the chiral components $\psi_+$ and $\psi_-$,  which follow from the mapping of higher-dimensional wave equations for half-integer spin in AdS space to the light-front~\cite{deTeramond:2013it, Brodsky:2014yha}
\begin{align} \label{LFDEa}  
- \frac{d}{d\zeta} \psi_-  - \frac{L+\half}{\zeta}\psi_-  -  V(\zeta) \psi_- &= M \psi_+ , \\  \label{LFDEb}  
 \frac{d}{d\zeta} \psi_+ - \frac{L+\half}{\zeta}\psi_+  - V(\zeta) \psi_+ &= M \psi_- , 
\end{align}
with normalization
\begin{align} \lb{norm}
 \int_0^\infty \! d \zeta \, \psi_+^2(\zeta) = \int_0^\infty \! d \zeta \, \psi_-^2(\zeta) = 1.
 \end{align}
The system of linear equations \req{LFDEa} and \req{LFDEb} is equivalent to the system of second order light-front wave equations
\begin{align}
 \Big(-\frac{d^2}{d\zeta^2} 
- \frac{1 - 4L^2}{4\zeta^2}+ U^+_\perp(\zeta) \Big)  \psi_+(\zeta) &= M_\perp^2 \psi_+(\zeta) ,   \\
 \Big(-\frac{d^2}{d\zeta^2} 
- \frac{1 - 4(L+1)^2}{4\zeta^2}+ U^-_\perp(\zeta) \Big)  \psi_-(\zeta) &= M_\perp^2 \psi_-(\zeta) , ~~
\end{align}
where
\begin{align} \label{Upm}
U^\pm(\zeta) = V^2(\zeta) \pm V'(z) + \frac{1 + 2 L}{\zeta} V(\zeta).
\end{align}
It corresponds to LF angular momentum $L$ and $L +1$.

\subsection{Invariant mass ansatz}

A simple ansatz to account for quark masses in HLFQCD was introduced in~\cite{Brodsky:2008pg} based on the arbitrary off-shell dependence of the LF wave function on the invariant mass squared: It controls the bound state according to the uncertainty principle in quantum mechanics. For a two-parton state this amounts to the substitution
\begin{align}
\frac{\mbf{k}_\perp^2}{x(1-x)} \to 
 \frac{\mbf{k}_\perp^2}{x(1-x)} + \frac{m_q^2}{x} + \frac{m_{\bar q}^2}{1-x},
 \end{align}
in the ground-state Gaussian wave function to include the expression for the LF kinetic energy with quark masses: It is also the invariant mass squared $s = (p_q + p_{\bar q})^2$ of the $q \bar q$ pair. This substitution leads, upon exponentiation, to  the longitudinal LF wave function~\cite{Brodsky:2008pg}
 \begin{align}  \label{IMWF}
 \chi_{IM}(x) =  \mathcal{N} \exp \Big( \! - \frac{1}{2 \lambda} \Big[\frac{m_q^2}{x} + \frac{m_{\bar q}^2}{1-x} \Big] \Big),
 \end{align}
and thus to a natural factorization of the longitudinal and transverse components of the LFWF, independent of the actual value of the effective quark masses. The label IM stands for invariant mass, and $\mathcal{N}$ is a normalization factor~\cite{Sandapen:2020nrn}.  On the other hand it was shown in~\cite{Gutsche:2012ez} that the ansatz
\begin{align}  \label{CSBWF}
 \chi_{CSB}(x) \sim x^a(1-x)^b,
 \end{align}
obtained in~\cite{tHooft:1974pnl, Bergknoff:1976xr}, leads to explicit charge symmetry breaking (CSB) in holographic QCD models.  The partonic mass shift contribution to hadron masses~\cite{Weisberger:1972hk},
\begin{align} \label{DeltaM2L}
\Delta M^2  =
 \int_0^1 dx \, \chi(x) \Big[ \frac{m_q^2}{x} +  \frac{m_{\bar q}^2}{1-x} \Big] \chi(x), 
\end{align}
used in~\cite{Brodsky:2008pg}  and~\cite{Gutsche:2012ez} does not incorporate, however,  the explicit contribution from a longitudinal potential to hadron masses required to satisfy the virial theorem.

\section{Longitudinal and transverse effective confinement potentials \lb{U}}

We briefly describe in this section the transverse confinement potential for mesons in HLFQCD and examine two particularly interesting models for longitudinal confinement:  the  well known 't Hooft  model~\cite{tHooft:1974pnl} in one-space and one-time directions, derived from first principles QCD in the large $N_C$ limit, and the effective potential model introduced by  Li, Maris, Zhao and Vary  (LMZV)~\cite{Li:2015zda}, which generates a complete basis function in the longitudinal direction.

\subsection{Transverse confinement potential \lb{TCP}}

The transverse LF equation \req{HT} has the same structure as the wave equations derived in  AdS  provided that one identifies $\zeta = z$~\cite{deTeramond:2008ht}, the holographic fifth-dimensional coordinate of AdS. This precise mapping allows us to relate the LF confinement potential $U_\perp$ to the dilaton profile which modifies AdS space~\cite{Brodsky:2014yha}. Conformal algebra underlies in LF holography the scale invariance of the QCD Lagrangian~\cite{Brodsky:2013ar}. It leads to the introduction of a scale $\lambda $ and harmonic confinement, $U \sim \lambda \zeta^2$,  maintaining the action conformal invariant~\cite{Brodsky:2013ar, deAlfaro:1976je}. The oscillator potential corresponds to a quadratic dilaton profile and thus to the emergence of linear Regge trajectories~\cite{Karch:2006pv}. 

Extension to superconformal algebra leads to a specific connection between mesons, baryons and tetraquarks~\cite{deTeramond:2014asa, Dosch:2015nwa, Brodsky:2016yod}  underlying the $SU(3)_C$ representation properties, since a diquark cluster can be in the same color representation as an antiquark, namely $\bar 3 \in 3 \times 3$.  The meson wave function $\phi_M$, the upper and lower components of the baryon wave function, $\phi_{B \,\pm} \equiv \psi_\pm$, and the tetraquark wave function, $\phi_T$, can be arranged as a supersymmetric 4-plet matrix~\cite{Brodsky:2016yod, Zou:2018eam}  
\begin{align}
\vert \Phi \rangle =   \begin{pmatrix}
    \phi_M^{\, (L+1)} & \phi_{B \, -}^{\, (L + 1)}\\
    \phi_{B \, + } ^{\, (L)} & \phi_T^{\, (L)}
    \end{pmatrix} ,
\end{align}
with $H_\perp^{LF} \vert \Phi \rangle = M^2_\perp \vert \Phi \rangle$. The constraints from superconformal structure uniquely determine the form of the effective transverse confining potential for mesons, nucleons and tetraquarks~\cite{deTeramond:2014asa, Dosch:2015nwa, Brodsky:2016yod}, including critical constant terms:
\begin{align} \lb{UMT}
U_{M \perp}(\zeta) &= \lambda^2 \zeta^2 + 2 \lambda (L_M - 1) ,\\
U^+_{B \perp}(\zeta) &= \lambda^2 \zeta^2 + 2 \lambda (L_B + 1) ,\\
U^-_{B \perp}(\zeta) &= \lambda^2 \zeta^2 + 2 \lambda L_B ,\\
U_{T \perp}(\zeta) &= \lambda^2 \zeta^2 + 2 \lambda (L_T + 1) ,
\end{align}
and lead to the remarkable relations  $L_M = L_B + 1$, $L_T = L_B$. The superconformal algebra also requires the universality of Regge slopes with a unique scale $\lambda$ for all hadron families, a well known fact in hadron physics, which holds to a good approximation.  

The solution of the AdS wave equations for mesons and vector mesons does incorporate the dependence on the total quark spin, $S = 0$ for the $\pi$ Regge trajectory and $S = 1$ for the $\rho$  trajectory: It is given by the additional term $2 \lambda S$, $S = 0, 1$, in the LF Hamiltonian.  It leads, for example to the correct prediction for the  $\pi - \rho$ mass gap: $M_\rho^2 - M^2_\pi = 2 \lambda$. In contrast, for baryons the quark spin interaction within a 2-quark cluster is not contained in the AdS action. To describe the quark spin-spin interaction, which distinguishes for example the nucleons from $\Delta$ particles, we also include an identical term, $2 \lambda S$, $S = 0, 1$ in the LF baryon Hamiltonian to maintain hadronic supersymmetry. The final expression for the transverse mass spectrum for mesons, baryons and tetraquarks is given by~\cite{Brodsky:2016yod}
\begin{align}
M^2_{M \perp} &= 4 \lambda (n+ L_M ) + 2 \lambda S,\\
M^2_{B \perp} &= 4 \lambda (n+ L_B+1) + 2 \lambda S,\\
M^2_{T \perp} & = 4 \lambda (n+ L_T+1) + 2 \lambda S,
\end{align}
with the same slope in $L$ and $n$, the radial quantum number.

\subsection{The t'Hooft model \lb{thooft}}

As discussed by Chabysheva and Hiller~\cite{Chabysheva:2012fe}, it is natural to identify the potential for longitudinal dynamics with the potential which underlies the ’t~Hooft model for large-$N_C$ QCD in (1+1) dimensions. It has the same form as the instantaneous LF potential which appears from instantaneous gluon exchange in the $A^+=0$ light-cone gauge in QCD (3+1)~\cite{Brodsky:1997de}.

In the 't Hooft model~\cite{tHooft:1974pnl}  the longitudinal equation~\req{HL} becomes  the integral equation for the pion
\begin{multline}  \lb{tHWE}
\Big( \frac{m_q^2}{x} +  \frac{m_{\bar q}^2}{1-x}\Big) \chi(x) + \frac {g^2 N_C}{\pi}   P \! \int_0^1 dx'  \frac{\chi(x) - \chi(x')}{(x - x')^2} \\ = M_\parallel^2  \, \chi(x) ,
\end{multline} 
where the coupling $g$ has dimensions of mass. To find an analytic solution to \req{tHWE} the approximate solution
\begin{align} \lb{chitH}
 \chi(x) \sim x^{\beta_1} (1-x)^{\beta_2} ,
 \end{align}
is chosen to cancel the end-point singularities~\cite{tHooft:1974pnl}.  Expanding \req{tHWE} near $x = \epsilon$ we find
\begin{align} \lb{TE}
\Big[\frac{\pi m_q^2}{g^2 N_C} - 1+ \pi \beta_1 \cot(\pi \beta_1)\Big] \epsilon^{\beta_1 - 1}  = 0.
\end{align}
It leads to the result   $\beta_1= (3 m_\pi^2/\pi g^2 N_C)^{1/2}$ from the expansion of the transcendental equation \req{TE}  for $m_q^2 / g^2 N_C \ll 1$. Likewise, we obtain   $\beta_2= (3 m_{\bar q}^2/\pi g^2 N_C)^{1/2}$ from the expansion of \req{tHWE}  at the upper bound $x = 1 - \epsilon$.

Integrating \req{tHWE} 
\begin{multline}  \lb{tHM2}
M_\parallel^2 =  \int_0^1 dx \, \Big( \frac{m_q^2}{x} +  \frac{m_{\bar q}^2}{1-x}\Big) \chi^2(x)  \\
+ \frac {g^2 N_C}{\pi}   P \! \int_0^1 dx \int_0^1 \! dx'  \, \frac{\chi(x)[\chi(x) - \chi(x’)]}{(x - x')^2} ,
\end{multline}
and using the approximate solution \req{chitH} in the limit of small quark masses we find for the pion mass
 \begin{align} \lb{chitH}
M_\pi^2 =  g \sqrt{\frac{\pi N_C}{ 3}} \, (m_q + m_{\bar q}) +  \mathcal{O} \! \left( (m_q \! + m_{\bar q})^2 \right),
 \end{align}
using the Hellmann-Feynman theorem to evaluate the second integral in \req{tHM2}. It has the same linear dependence in the quark mass as the Gell-Mann-Oakes-Renner  (GMOR) relation~\cite{GellMann:1968rz}. Spontaneous chiral symmetry breaking in the  't Hooft model occurs in the limit $N_C \to \infty$ followed by the limit $m_q \to 0$~\cite{tHooft:1974pnl, Zhitnitsky:1985um}.

\subsection{The LZMV longitudinal potential \lb{LZMV}}

The effective LMZV potential~\cite{Li:2015zda} 
\begin{align} \label{UL}
U_\parallel(x) =  - \sigma^2 \frac{d}{d x}  \Big(x (1-x) \, \frac{d}{d x} \Big)
\end{align}
was introduced in the context of basis light-front quantization (BLFQ)~\cite{Vary:2009gt, Li:2013cga}. It generates a complete basis of orthonormal functions from the solution of \req{HL}
\begin{align}  \label{chikappa}
\chi^{\alpha, \beta}_\kappa(x) = N  x^{\alpha/2}  (1-x)^{\beta/2}  P_\kappa^{(\alpha, \beta)}(1- 2x),
\end{align}
where $\alpha = 2 m_q/\sigma$, $\beta = 2 m_{\bar q}/\sigma$,  $ P_\kappa^{(\alpha, \beta)}$ is a Jacobi polynomial of order $\kappa$ and $N$ is a normalization factor (Appendix \ref{A}). The longitudinal basis can be combined with the holographic LFWF basis in transverse space~\cite{Brodsky:2014yha} to perform extensive numerical computations using the BLFQ methods~\cite{Li:2017mlw, Xu:2019xhk, Qian:2020utg}, as well  other computations which require a complete basis function~\cite{Sheckler:2020fbt}, including numerical evaluation in quantum computers~\cite{Kreshchuk:2020aiq}.

The potential \req{UL}  leads to an oscillator  in the longitudinal dimension which is combined with the transverse oscillator for heavy quark masses  $m_q, m_{\bar q} \to  m_Q, m_{\overline{Q}}$~\cite{Li:2015zda}.  The recovery of the rotational invariance of the oscillator in the nonrelativistic limit leads to a precise connection of the dimensionful longitudinal coupling $\sigma$ to the transverse scale $\lambda$ (Appendix \ref{B}) 
\begin{align} \label{siglamb}
\sigma = \frac{\lambda_Q}{m_Q + m_{\overline{Q}}} ,
\end{align}
where $\lambda_Q$ depends on the heavy quark mass $m_Q$, $\lambda \to \lambda_Q$, to keep the ratio \req{siglamb} constant in the heavy quark mass domain (See \ref{MQ}). More relevant for our present purposes, in addition to its simplicity,  is the observation that in the limit of small quark masses, the potential \req{UL} leads to the approximate solution \req{chitH} obtained in~\cite{tHooft:1974pnl, Bergknoff:1976xr}, thereby incorporating the breaking of chiral symmetry in the LF holographic model consistent with the virial theorem.

\section{Extended HLFQCD model \lb{LFEM}}

The extension of HLFQCD to include quark masses described in this article combines the transverse and longitudinal dynamics as in~\cite{Chabysheva:2012fe, Li:2015zda}, but  keeping the physically motivated invariant mass ansatz, which is required in order to incorporate the invariant off-mass-shell dependence of the full LFWF as in our previous work. In practice, we include the longitudinal dynamics by performing an expansion of the invariant mass LFWF  \req{IMWF} using the convenient basis of complete orthonormal eigenfunctions generated by the specific potential \req{UL}
\begin{align} \label{BE} \nonumber
\chi_{IM}(x) &= \mathcal{N } \exp  \Big( - \frac{\sigma^2}{8 \lambda} \Big[\frac{\alpha^2}{x} + \frac{\beta^2}{1-x} \Big] \Big) \\
&= \sum_\kappa C_\kappa \,  \chi_\kappa(x) .
 \end{align}
 The expansion coefficients $ C_\kappa$ are computed from the overlap
 \begin{align}
 C_\kappa^{\alpha, \beta}  = \int_0^1 dx \, \chi_\kappa^{\alpha, \beta}(x)  \chi_{IM}(x),
 \end{align}
 with $\langle  \chi_{IM} \vert  \chi_{IM} \rangle  = \sum_\kappa C_\kappa^2 = 1$.
A rapid convergence of the invariant mass LFWF was obtained in~\cite{Chabysheva:2012fe, Li:2015zda} using the basis wave function \req{chikappa}, with the lowest-index Jacobi polynomial giving the dominant contribution.

In terms of  \req{chikappa} the longitudinal contribution to a meson mass is given by
\begin{multline} \label{M2Lsigma}
M_\parallel^2  / \sigma^2 =
   \int_0^1\! dx  \, \chi_{IM}(x) \Big(- \frac{d}{d x}  \Big(x (1-x) \, \frac{d}{d x} \Big) 
   \\ + \frac{1}{4} \Big[\frac{\alpha^2}{x} +  \frac{\beta^2}{1-x} \Big] \Big) \chi_{IM}(x)  \\
=  \sum_\kappa C_\kappa^2 \, \nu^2 (\kappa, \alpha, \beta), \hspace{90pt}
\end{multline}
expressed of a sum of the eigenvalues 
\begin{align}
\nu^2 (\kappa, \alpha, \beta) =  \fourth  (\alpha + \beta + 2 \kappa) (2 + \alpha + \beta + 2 \kappa),
\end{align}
computed in Appendix \ref{A}. In terms of the constituent quark masses the longitudinal mode expansion \req{M2Lsigma} can be expressed as
\begin{multline} \lb{M2par}
M_\parallel^2 = \left(m_q + m_{\bar q}\right)^2 +   \sigma  \left(m_q + m_{\bar q}\right) \\
+ \sum_{\kappa = 1}^\infty C_\kappa^2 \left[ (2 \kappa  \sigma  \left(m_q + m_{\bar q}\right)  + \kappa (\kappa + 1) \sigma^2 \right].
\end{multline}
The rapid convergence of \req{M2par} in the basis function \req{chikappa} found in~\cite{Chabysheva:2012fe, Li:2015zda} and discussed below in Sec.~\ref{NR}, effectively ensures that the longitudinal mass squared does not grow as $\kappa^2$ for large $\kappa$, which would contradict linear Regge behavior.

\subsection{Chiral symmetry breaking}

The chiral limit follows directly from  \req{M2par}  since all the coefficients $C_\kappa$ vanish for $\kappa \ne 0$ in this limit (see \req{chi00}). We obtain for the pion mass
\begin{align} \label{chi}
M^2_\pi =   \sigma (m_u \! + m_d) +   \mathcal{O} \! \left( (m_u \! + m_d)^2 \right) ,
\end{align}
in the limit $m_u, m_d \to 0$. It has the same linear GMOR dependence in the quark mass~\cite{GellMann:1968rz} 
\begin{equation}
M^2_\pi f_\pi^2 = - \half (m_u \! + m_d) \langle \bar u u + \bar d d \rangle  + \mathcal{O} \! \left( (m_u \! + m_d)^2 \right).
\end{equation}
Formally the longitudinal coupling $\sigma$ sigma is given by $\sigma = - \langle \bar \psi  \psi \rangle/ f_\pi^2$, where the vacuum condensate $\langle \overline \psi \psi \rangle \equiv  \half \langle \bar u u + \bar d d \rangle$ plays the role of a chiral order parameter.  The same linear dependence in \req{chi} arises for the  (3 + 1) effective  LF Hamiltonian, since the constraints from the superconformal algebra require that the contribution to the pion mass from the transverse LF dynamics is identically zero~\cite{Dosch:2015nwa}. 

Comparison with \req{chitH} leads to  
\begin{align} \label{sigmagNC}
\sigma =  2 g \sqrt{\frac{\pi N_C}{3}}  = \text{const},
\end{align}
since $g$ scales as $g \sim 1 / \sqrt{N_C}$ and chiral logarithms are suppressed at $N_C \to \infty$. Both \req{chi} and \req{chitH} receive identical contributions from the potential and kinetic energy terms in agreement with the virial theorem.

\subsection{Heavy quark masses \lb{MQ}}

\begin{figure}[h] 
\includegraphics[width=8.6cm]{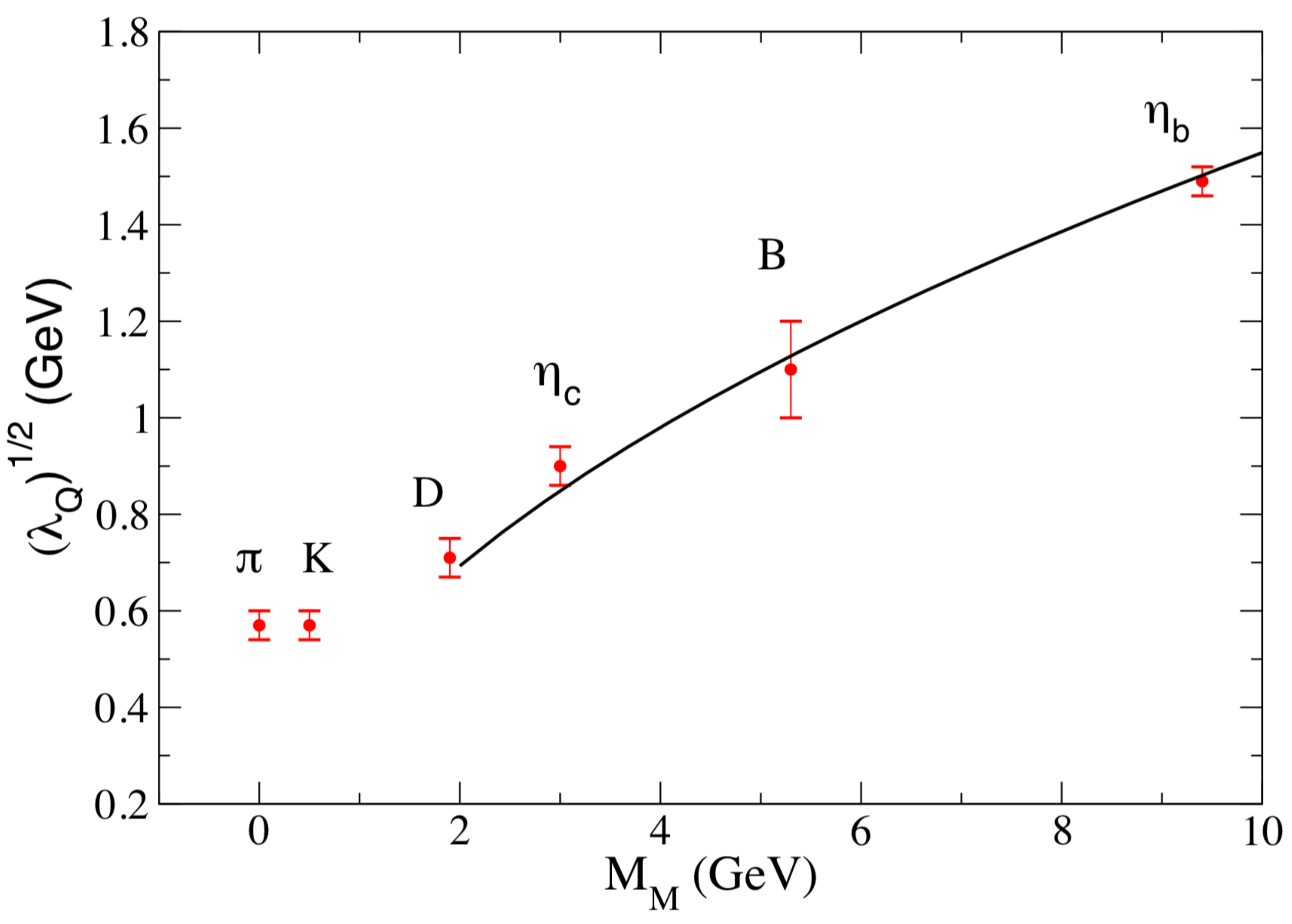}
\caption{\label{lambda}  Fitted values of $\sqrt{\lambda_Q}$ as a function of the mass of the lowest meson state $M_M$ on the Regge trajectory from Ref~\cite{Nielsen:2018ytt}. The solid line is the fit from Eq.~\req{laQMQ}.}
\end{figure}

It is well known~\cite{Shuryak:1981fza} that for heavy mesons the product $\sqrt M f_M$, where $f_M$ is the meson decay constant, approaches, up to logarithmic terms, a finite value $\sqrt M f_M \to  C$, in agreement with the heavy-quark effective theory (HQET) result~\cite{Isgur:1991wq}. In the LF holographic context this means that the confinement scale $\lambda_Q$ has to increase with increasing quark mass $m_Q$. In the limit of heavy quarks the meson mass equals the sum of quark masses ({\it cf.}~Eq.~\req{M2par}),  which requires that the confining scale is proportional to the mass of the heavy meson $M_M$~\cite{Dosch:2016zdv, Gutsche:2012ez}
\begin{align} \lb{laQMQ}
\sqrt{\lambda_Q} = C \sqrt{M_Q}.
\end{align}
In~\cite{Nielsen:2018ytt} the value  $C = 0.49 \pm 0.02 ~ \text{GeV}^{1/2}$ was found by fitting \req{laQMQ} in the mass interval  $2 \le M_M \le 10$ GeV as shown in Fig. \ref{lambda}. Thus 
\begin{align} \label{sigMN}
\sigma \simeq \frac{\lambda_Q}{M_M} \simeq C^2  \simeq  0.24 ~ {\rm GeV}.
\end{align}
We fix the longitudinal coupling $\sigma$ to the value found in~\cite{Nielsen:2018ytt}, namely $\sigma =   0.24$ GeV.  We have kept the same mass dependence of the transverse scale $\lambda$  (Fig. \ref{lambda}) since it reproduces quite well the slopes of Regge families across the hadron spectrum.    We have extrapolated the value of $\sigma$ to  the light quark mass domain, where we have no guidance from HQET, by assuming that $\sigma$ remain approximately constant, a result supported by the large $N_C$ result \req{sigmagNC} and consistency with the chiral limit.

\section{Numerical results \lb{NR} }

Having fixed the value of the longitudinal confinement scale $\sigma$ by the rate of change of the transverse confinement scale $\lambda$, we need to determine the value of the effective quark masses in \req{M2par} to compute the longitudinal mass contribution $M_\parallel^2$ to the total hadron mass. Thus, we determine the effective light quark masses $m_u$ and $m_d$ from the measured pion mass and  the strange quark mass, $m_s$, from the kaon mass using \req{M2par}: The value of the $\phi(1020)$ mass is then a prediction. Notice that the $\phi(1020)$ vector meson also has the transverse mass component $M_\perp = \sqrt{2 \lambda}$ from the spin-spin interaction in supersymmetric LF holographic QCD~\cite{Brodsky:2014yha, Brodsky:2016yod} with $\sqrt{\lambda}= 0.523 \pm 0.024$ GeV.

\begin{table}[htp] 
\caption{Lowest order coefficients $C_\kappa$  in the expansion \req{BE}.}
\begin{center}
\begin{tabular}{|c| c c c c c c  c|}
\hline \hline
 &  $\kappa = 0$ ~& $\kappa = 1$~& $\kappa = 2$ ~ & $\kappa = 3$ ~ &  $\kappa = 4$ ~ & $\kappa = 5$ ~ & $\kappa = 6$\\ 
 \hline  
  $C(u \bar d)$ & 0.998 &0 & 0.055 & 0 & 0.010 & 0 & -0.003 ~\\ 
 $ C(u \bar s)$ &  0.967 &-0.231 & 0.100 & -0.006 & -0.009 & 0.013 & -0.016 \\
 $ C(s \bar s)$ & 0.998 & 0  & 0.038 & 0 & -0.045 & 0 & -0.024 \\
 $ C(u \bar c)$ & 0.958 & -0.267 & 0.097& -0.012 &  -0.003&  0& -0.007  \\
 $ C(c \bar c)$ & 0.999 & 0  & 0.016 & 0 & -0.020 & 0 &  -0.003 \\
 \hline    \hline
\end{tabular}
\end{center}
\label{Ckappa}
\end{table}

\begin{figure}[htp]
\begin{center}
\includegraphics[width=8.6cm]{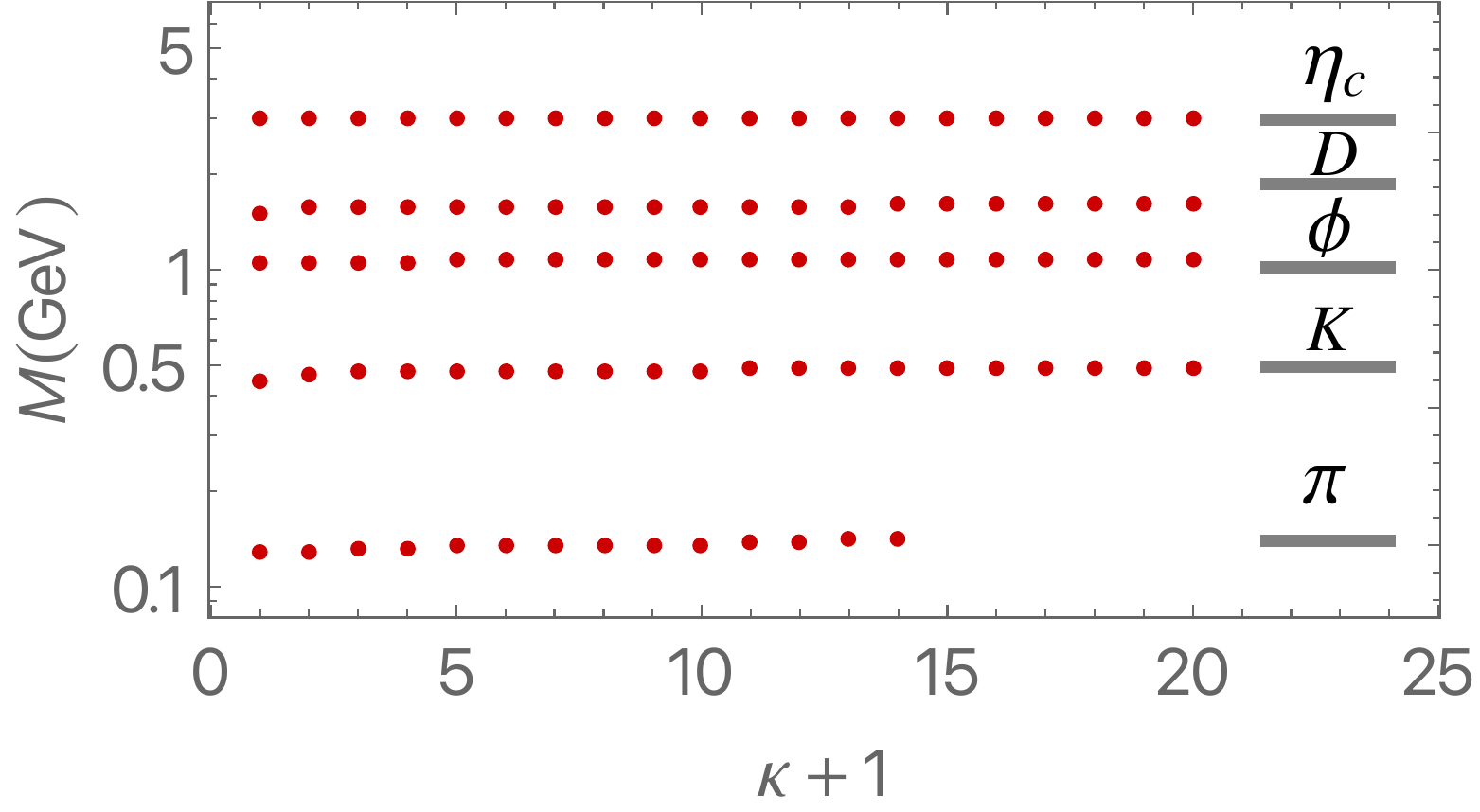}
\end{center}
\caption{\label{LMMs}Numerical evaluation of ground state meson masses from \req{M2Lsigma}. The horizontal grey lines in the figure are the observed masses~\cite{Zyla:2020zbs} also given in  Table \ref{PDGcomp}  for comparison.}
\end{figure}

We show in Table~\ref{Ckappa} the  lowest  order coefficients  in the expansion \req{BE}. The results for the light meson masses  in Fig.~\ref{LMMs} correspond to the values $m_u = m_d = 28 $ MeV and  $m_s =   326  $ MeV.  The quark masses $m_q$ and $m_{\bar q}$  in \req{HL} are effective  quark masses  from the renormalization due to the reduction of higher Fock states as functionals of the valence state~\cite{Pauli:1998tf},  not the current quark masses in the QCD Lagrangian.  The actual value of $\sigma$ used here is determined from the heavy quark mass constraint \req{siglamb} to have a continuous description across the light and heavy scales.  If instead $\sigma$ is determined by the current quark masses one obtains the value  $\sigma \simeq 3 ~{\rm GeV}$, namely an order of magnitude larger than the transverse scale $\sqrt{\lambda} \simeq 0.5 ~{\rm GeV}$, yielding an unphysical spectrum.

\begin{table}[htp] 
\caption{Comparison of meson mass results in Fig.~\ref{LMMs} (2nd column) with observed values from~\cite{Zyla:2020zbs} (3rd column). One-gluon exchange corrections are not included for the charmonium and bottonium.}
\begin{center}
\begin{tabular}{|c| c  c|}
\hline \hline
 & $M (\rm{GeV})$  & ~~$M_{\rm data}$ (\rm{GeV}) \\
 \hline  
  $\pi^{\pm}$  &  0.140&  0.140 \\ 
 $ K^\pm $ &  0.490 &  0.494 \\
 $ \phi(1020)$ & 1.067 &  1.019  \\
 $ D^\pm$ & 1.602 &  1.870   \\
 $ \eta_c(1S)$ & 2.978 &  2.984   \\
 \hline    \hline
\end{tabular}
\end{center}
\label{PDGcomp}
\end{table}

\begin{figure}[htp]
\begin{center}
\includegraphics[width=8.6cm]{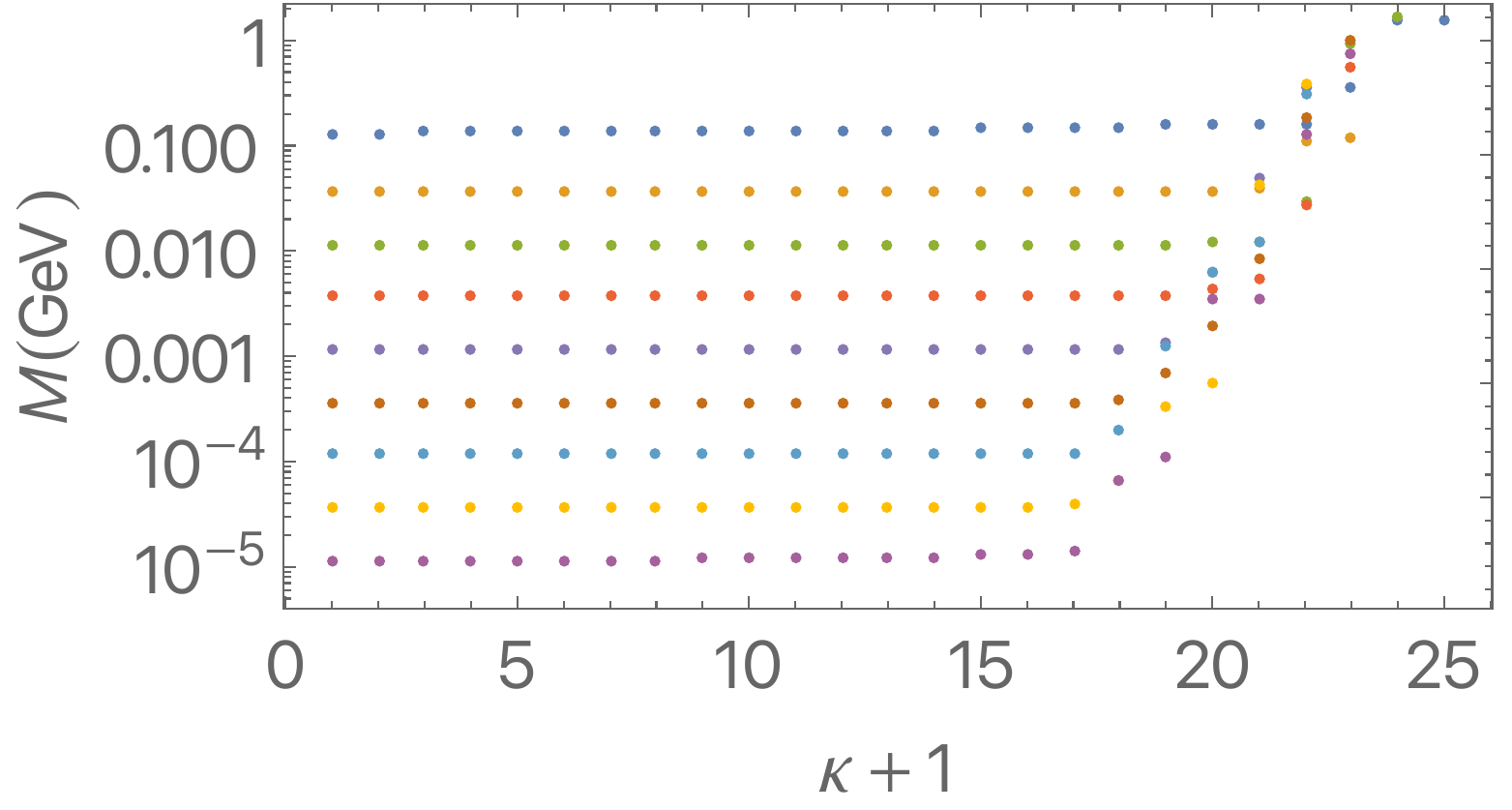}
\end{center}
\caption{\label{CML}  The numerical evaluation of the pion mass for quark masses in the range $m_q = 28$ MeV (upper blue dotted line) to $m_q = 28 \times 10^{-8}$ MeV (lower mauve dotted line), for fixed $\sigma$, manifests the numerical spurious divergence from highly oscillatory integrands at large  $\kappa$, in sharp contrast with the exact chiral result \req{chi}.}
\end{figure}

Meson masses are determined from the stability plateau in Fig.~\ref{LMMs}.  For light quark masses, contributions above $\kappa_{max} \simeq 20$ introduce large uncertainties  from highly oscillatory integrands.  In Fig. \ref{CML} we show the effect of the strong oscillations of the Jacobi Polynomials at large $\kappa$~\cite{Szego} by examining the dependence of the pion mass, $M_\pi$, for quark masses in the interval $m_q = 28$ MeV to $m_q = 28 \times 10^{-8}$ MeV for fixed $\sigma$.

The meson distribution amplitudes (DAs)~\cite{Lepage:1979zb}
\begin{align}
\phi_M(x) \sim X(x) = \sqrt{x(1-x)} \, \chi(x),
\end{align}
for the pion, kaon, D and $J/ \Psi$ mesons are shown in Fig.~\req{LX}. Due to the rapid convergence of the exponential wave function in the basis expansion \req{BE}, very few modes are required to reproduce the invariant mass LFWF. The DAs predicted by HLFQCD at the initial nonperturbative scale should then be evolved to the relevant scale using the ERBL equation~\cite{Lepage:1979zb, Efremov:1979qk, Brodsky:2011yv}. The Dyson-Schwinger results for the pion DA~\cite{Roberts:2021nhw} are very similar to the chiral limit result $\phi_\pi(x) \sim \sqrt{x(1-x)}$ from  LF holographic mapping~\cite{Brodsky:2006uqa}.

\begin{figure}[htp]
\begin{center}
\includegraphics[width=4.276cm]{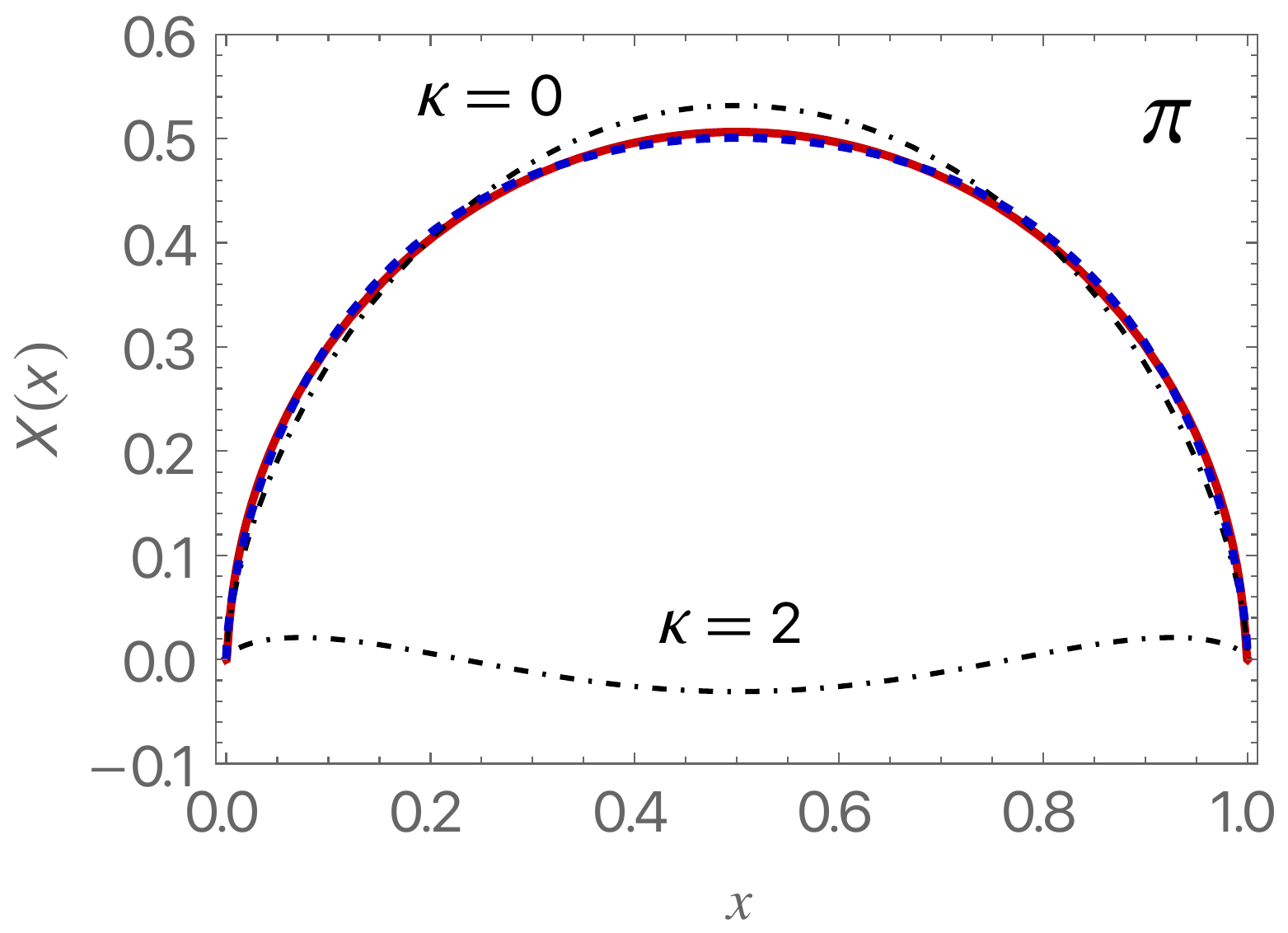}
\includegraphics[width=4.272cm]{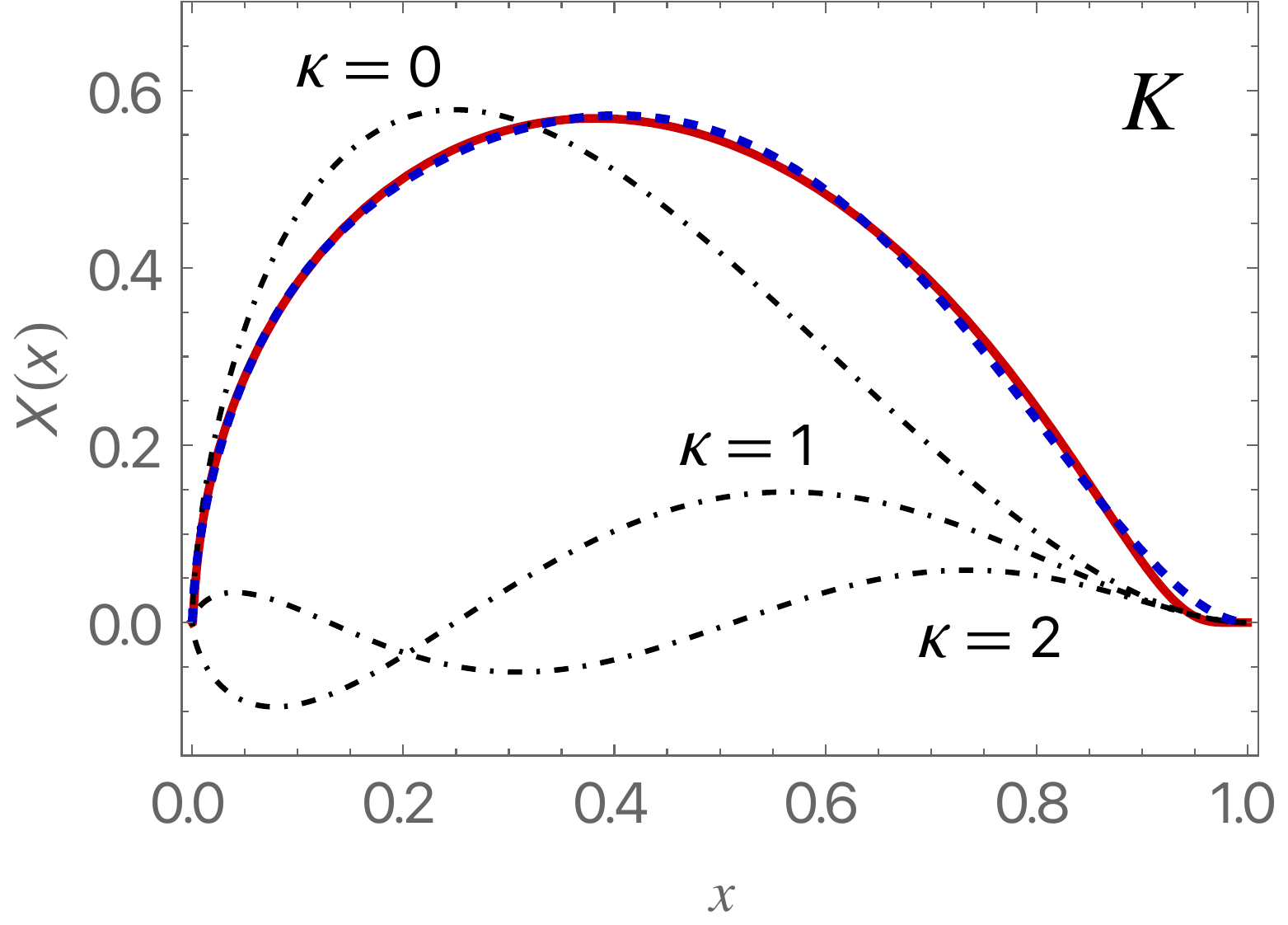}
\includegraphics[width=4.275cm]{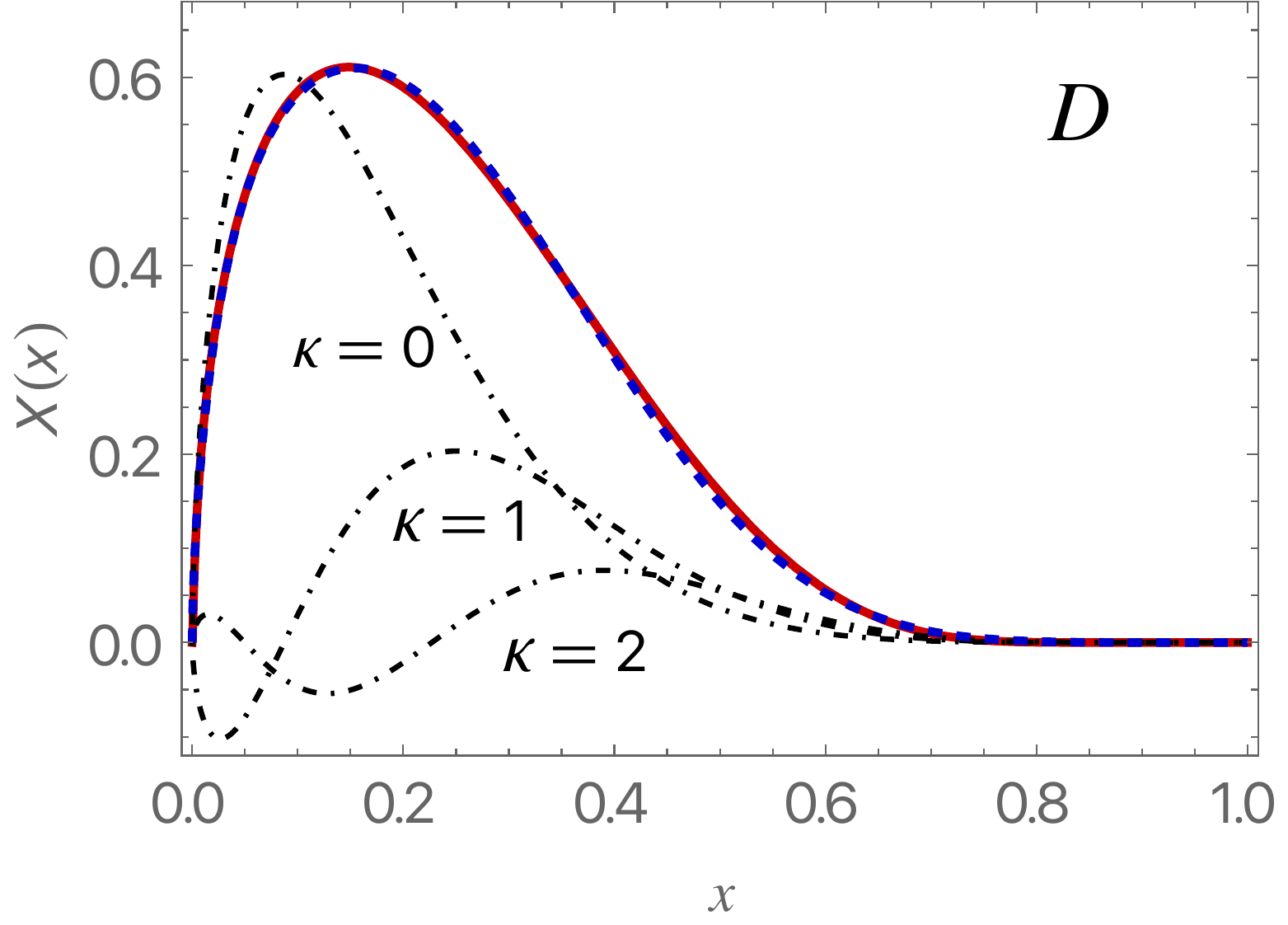}
\includegraphics[width=4.275cm]{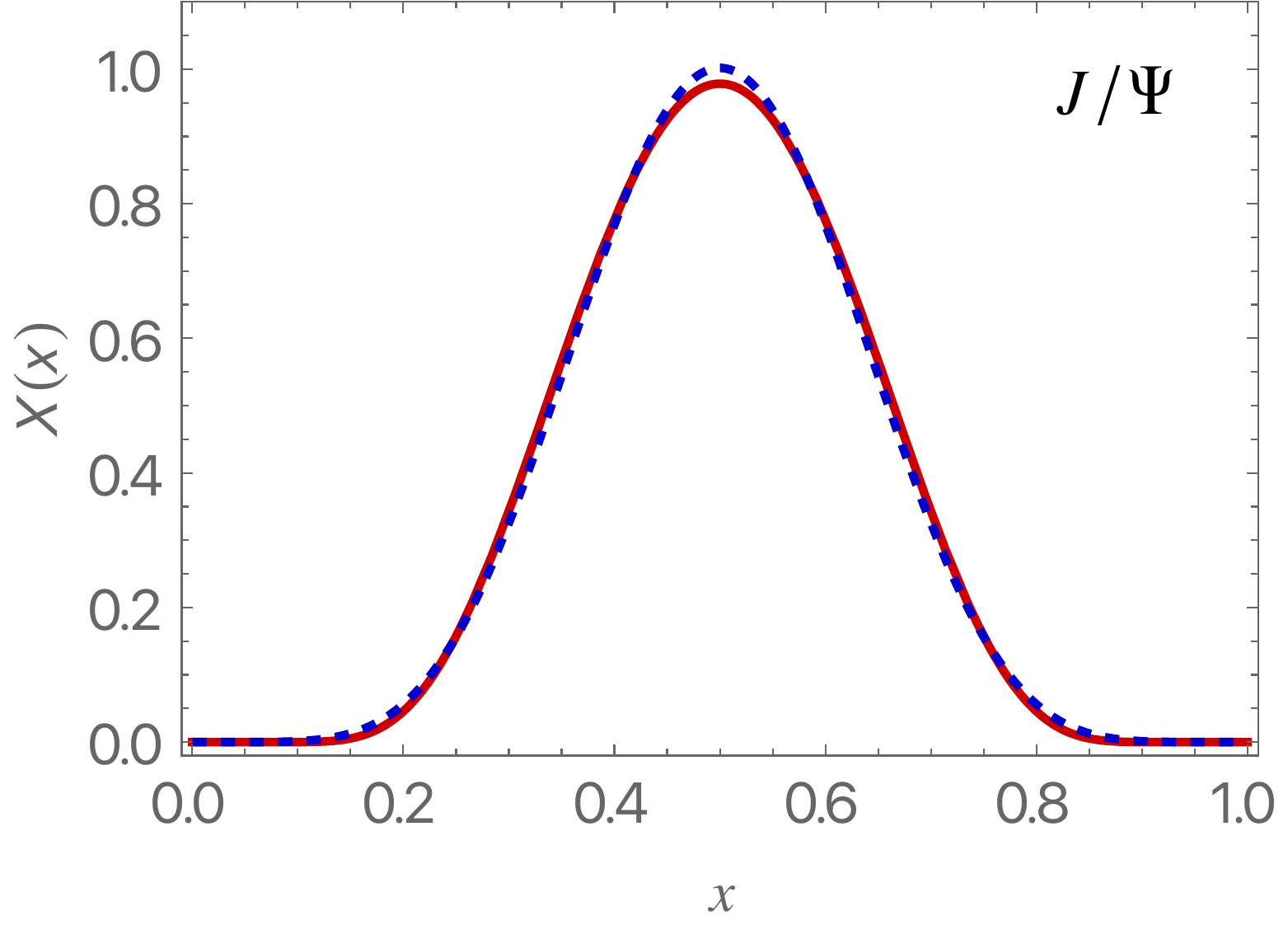}
\end{center}
\caption{\label{LX} Light-front distribution amplitudes $X(x)$ for the $\pi$, $K$, $D$ and $J/ \Psi$ mesons: the red curve is the invariant mass result,  dot dashed black curves are individual modes in the expansion \req{BE}, dashed blue curve represent the sum of  modes in the figure. Notice that the $J/\Psi$ result is well described by the zero-order Jacobi Polynomial.}
\end{figure}

Finally, we can extend our analysis to the heavy quark sector provided that longitudinal and transverse dynamics can be separated to a good approximation. In contrast with  light quark masses $m_q, m_{\bar q} \ll \sigma$, $q = u, d$, most of the hadron mass in the heavy sector, $m_Q, m_{\bar Q} \gg \sigma$, comes from  quark masses.  The expansion coefficients for the wave function  \req{BE}  for the $u c$  and $c c$ mesons are shown in Table \ref{Ckappa}. We determine the effective charm quark mass from the $\eta_c$ using \req{M2par} and compute, for example, the mass of the $D$ meson as a prediction. We find for $M_D$  a value within 14\% of its measured value for $m_c \simeq 1.4$ GeV (Table \ref{PDGcomp}). Our results do not include the negative contribution from one-gluon exchange for the small size double-heavy mesons~\cite{Gutsche:2012ez, Li:2015zda} which gives a larger value for the extracted charmed quark mass, therefore a value for the $D$ meson mass closer to its experimental value.

\section{Conclusions and Outlook \lb{CaO}}

The light-front semiclassical approximation described in this article incorporates the confinement strength in the longitudinal direction as well as the effective scale of chiral symmetry breaking as different manifestations of the same underlying dynamics.  As such, it accounts for most of the meson mass, consistent with the GMOR relation. Following~\cite{Chabysheva:2012fe} we have separated LF longitudinal and transverse dynamics and adopted from~\cite{Li:2015zda} a potential which generates a convenient basis function in the longitudinal direction. It is responsible for breaking the chiral symmetry and restoring rotational symmetry in the limit of heavy quark masses.  In practice, it allows us to reduce the highly complex and non-local four-dimensional LF QCD Hamiltonian to a set of two independent second order differential equations, consistent with the local structure of semiclassical AdS equations~\cite{Narayanan:2005gh, Katz:2007br}.

The mass dependence of the transverse scale $\lambda$ found in \cite{Dosch:2016zdv, Nielsen:2018ytt} is unchanged, therefore the Regge slopes are unmodified. The changes introduced by the new approach primarily modify the lowest meson state in each Regge trajectory, and therefore the value of the extracted effective quark masses. Our previous successful results for the Regge trajectories for light and heavy quark masses are basically unchanged.
 
The origin and physical interpretation of the longitudinal scale $\sigma$, which has the role of a condensate $\langle \bar \psi \psi \rangle$, remains to be explored,  but as we have shown in \req{sigmagNC}, it is  related to the dimensionful constant $g$ in QCD$(1 + 1)$ at large $N_C$.  In  lattice QCD the structure of the vacuum is sampled in the Euclidean region where non-trivial gauge field configurations  provide a mechanism for symmetry breaking through the Banks-Casher relation, $\langle \bar \psi \psi \rangle = - \pi \rho(0)$, with $\rho(0)$  the density or Dirac-zero modes~\cite{Banks:1979yr, McNeile:2012xh}. However, the relation between chiral symmetry breaking and confinement has remained elusive. In this context, it has been argued that the chiral condensate, usually viewed as a constant mass scale which fills all spacetime, is instead contained within hadrons, therefore a property of hadron dynamics~\cite{Brodsky:2012ku, Mannheim:2019lss}.

The fact that the nonzero pion mass is a consequence of longitudinal LF confinement is a remarkable result. One would expect, for example, from the two-dimensional 't~Hooft~\cite{tHooft:1974pnl} or Schwinger~\cite{Bergknoff:1976xr} models in light-front coordinates, that  the pion mass from the mechanism of chiral symmetry breaking originates in the longitudinal component of the wave function~\cite{Zhitnitsky:1985um}, since the kinetic quark mass terms only depend on the longitudinal variable. However, this would not be the case if the transverse kinetic and potential energy of the pion would not exactly cancel as required by the superconformal structure of the transverse LF Hamiltonian.   The pion plays a special role as a hadronic state of zero mass in the chiral limit. Since it does not have a baryonic partner, the pion breaks the meson-baryon hadron supersymmetry~\cite{Dosch:2015nwa}.  In contrast, the proton mass (as well as the mass of radial and orbital hadron excited states) is generated by the addition of the transverse kinetic and transverse potential energy with a small contribution from the longitudinal dynamics, in agreement with the Regge phenomenology of the hadron mass spectrum.

\acknowledgments

We thank Yang Li and James Vary for their collaboration in the early stages of this work and for reading our manuscript. We also thank Hans Guenter Dosch for critical remarks and Alexandre Deur, John Hiller, Valery Lyubovitsky, Gerald Miller and Ruben Sandapen for useful comments. The work of SJB is supported in part by the Department of Energy, Contract DE--AC02--76SF00515, SLAC-PUB-17593.

\subsection*{Note added in proof}

We refer the reader to the related work by Y.~Li and J.~P.~Vary, Light-front holography with chiral symmetry breaking, arXiv:2103.09993 [hep-ph] submitted concurrently with this article, as well as to M.~Ahmady, H.~Dahiya, S.~Kaur, C.~Mondal, R.~Sandapen and N.~Sharma, Extending light-front holographic QCD using the 't Hooft Equation,
Phys. Lett. B \textbf{823}, 136754 (2021) [arXiv:2105.01018 [hep-ph]] and to M.~Ahmady, S.~Kaur, S.~L.~MacKay, C.~Mondal and R.~Sandapen, Hadron spectroscopy using the light-front holographic Schr\"odinger equation and the \textquoteright{}t Hooft equation, Phys. Rev. D \textbf{104},  074013 (2021) [arXiv:2108.03482 [hep-ph]], submitted after ours. We thank Colin Weller and Gerald Miller for sharing with us their  related upcoming article on confinement in two-dimensional QCD.

\appendix

\section{Jacobi polynomials and solution to the longitudinal Hamiltonian equation \lb{A}}

The Jacobi polynomials $P_\kappa^{(\alpha, \beta)}(z)$ are  solution of the differential equation
\begin{multline} \label{JP}
(1-z)^{-\alpha} (1+z)^{-\beta} \frac{d}{d z} \Big( (1-z)^{\alpha+1} (1+z)^{\beta+1}  \frac{d}{d z} \, u(z) \Big)  \\
+ \kappa(\kappa+ a + b + 1) u(z) = 0, 
\end{multline}
which is orthogonal in the  interval  $[-1, 1]$ with weight $(1-z)^\alpha  (1-z)^\beta$. Performing the change of variable $z = 1 - 2 x$ we find
\begin{multline}
x^{-\alpha} (1-x)^{-\beta}  \frac{d}{d x} \Big( x^{\alpha+1} (1- x)^{\beta+1}  \frac{d}{d x}  \, u(x) \Big)  \\
+ \kappa(\kappa+ a + b + 1) u(x) = 0,
\end{multline}
with the solution $P_\kappa^{(\alpha, \beta)} (1 - 2 x) $  orthogonal in the interval  $[0, 1]$ with weight $x^\alpha (1-x)^\beta$.

Consider now the eigenvalue equation
 \begin{equation} \lb{eigneq}
\left( -  \frac{d}{d x}  \Big(x (1-x) \, \frac{d}{d x} \Big)
 + \frac{1}{4} \Big[\frac{\alpha^2}{x} +  \frac{\beta^2}{1-x} \Big]\right) v(x) =  \nu^2 v(x) .
\end{equation}
Writing $v(x) = x^{\alpha/2}  (1-x)x^{\beta/2} w(x)$ and substituting in \req{eigneq}  we find  that $w(x) =  P_\kappa^{(\alpha, \beta)} (1 - 2 x)$. 
Therefore the normalized solution to \req{eigneq} 
\begin{align} \lb{chiab}
\chi^{\alpha, \beta}_\kappa(x) = N  x^{\alpha/2}  (1-x)^{\beta/2}  P_\kappa^{(\alpha, \beta)}(1- 2x),
\end{align}
with eigenvalues
\begin{align}
\nu^2 = \fourth (\alpha + \beta + 2 \kappa) (2 + \alpha + \beta + 2 \kappa),
\end{align}
and normalization
\begin{equation}
N = \sqrt{1 + \alpha + \beta + 2 \kappa}  \,
\sqrt{\frac{ \Gamma(1 +  \kappa) \Gamma(1 + \alpha + \beta + \kappa)}{\Gamma(1 + \alpha + \kappa) \Gamma(1 + \beta + \kappa)}},
\end{equation}
given in terms of the Gamma function. If $\alpha = \beta = 0$ the solution \req{chiab} is reduced to
 \begin{align} \lb{chi00}
\chi^{0, 0}_\kappa(x) = \sqrt{1 + 2 \kappa} \, P_\kappa(1 - 2x),
\end{align}
where $P_\kappa(x)$ is a Legendre polynomial.

\vspace{15pt}

\section{Heavy quark mass limit \lb{B}}

Rotational invariance of the oscillator potential in the non relativistic limit leads to a precise connection of the scale    $\sigma$, of the longitudinal potential in Sec.~\ref{LZMV},  with the strength $\lambda$ of the transverse confinement potential in Sec.~\ref{TCP}, which sets the hadronic scale. To show this, consider the nonrelativsitic limit 
$m_q, m_{\bar q} \to  m_Q, m_{\bar Q} \gg k_\perp,  k_z$, $\lambda \to \lambda_Q$ with  
\begin{align}
x = \frac{m_Q + k_z}{m_Q + m_{\overline Q}}, \quad \
\overline x = \frac{m_{\overline Q} - k_z}{m_Q + m_{\overline Q}}. 
\end{align}
One finds 
the rotationally-invariant potential
\begin{align}
U(r)  \to V(r)  \equiv \frac{U(r)}{m_Q + m_{\overline Q}}  = \half \mu \, \omega^2 r^2,
\end{align}
and the constraint
\begin{align} \label{sigomlamb}
\omega = \sigma = \frac{\lambda_Q}{m_Q + m_{\overline Q}} ,
\end{align}
in the limit of heavy quark masses where
\begin{align}
\mu =  \frac{m_Q m_{\bar Q}}{m_Q + m_{\overline Q}},
\end{align}
is the reduced mass of the heavy quark-antiquark system and
$\mbf{r}^2 = \mbf{b}_\perp^2 + b_z^2$~\cite{Miller:2019ysh}.


\end{document}